\documentclass[twocolumn,amsmath,amssymb,superscriptaddress,prl,nobibnotes]{revtex4-1}

\usepackage{graphicx}

\usepackage{amsmath}
\usepackage{amssymb}
\usepackage{dcolumn}
\usepackage{bm}
\usepackage{upgreek}

\DeclareMathAlphabet\mathbfcal{OMS}{cmsy}{b}{n}
\usepackage{color}

\begin{document}

\title{Laser Amplification in Strongly-Magnetized Plasma}

\author{Matthew R. Edwards}
\thanks{M.R. Edwards and Y. Shi contributed equally to this work.}
\affiliation{Department of Mechanical and Aerospace Engineering, Princeton University, Princeton, New Jersey 08544, USA}%
\author{}
\email{mredward@princeton.edu}
\noaffiliation{}
\author{Yuan Shi}
\thanks{M.R. Edwards and Y. Shi contributed equally to this work.}
\affiliation{Department of Astrophysical Sciences, Princeton University, Princeton, New Jersey 08544, USA}%
\affiliation{Princeton Plasma Physics Laboratory, Princeton, New Jersey 08543, USA}
\affiliation{Lawrence Livermore National Laboratory, Livermore, California 94550, USA}
\author{Julia M. Mikhailova}
\email{j.mikhailova@princeton.edu}
\affiliation{Department of Mechanical and Aerospace Engineering, Princeton University, Princeton, New Jersey 08544, USA}%
\author{Nathaniel J. Fisch}
\email{fisch@princeton.edu}
\affiliation{Department of Astrophysical Sciences, Princeton University, Princeton, New Jersey 08544, USA}%
\affiliation{Princeton Plasma Physics Laboratory, Princeton, New Jersey 08543, USA}

\date{\today}

\begin{abstract}
We consider backscattering of laser pulses in strongly-magnetized plasma mediated by kinetic magnetohydrodynamic waves. Magnetized low-frequency scattering, which can occur when the external magnetic field is neither perpendicular nor parallel to the laser propagation direction, provides an instability growth rate higher than Raman scattering and a frequency downshift comparable to Brillouin scattering. In addition to the high growth rate, which allows smaller plasmas, and the 0.1-2\% frequency downshift, which permits a wide range of pump sources, MLF scattering is an ideal candidate for amplification because the process supports an extremely large bandwidth, which particle-in-cell simulations show produces ultrashort durations. Under some conditions, MLF scattering also becomes the dominant spontaneous backscatter instability, with implications for magnetized laser-confinement experiments. 
\end{abstract}

\date{\today}
\maketitle

Laser-driven magnetic field generation has produced kilotesla (10 MG) field strengths with coil-type targets \cite{Knauer2010compressing,Yoneda2012strong,Fujioka2013kilotesla,Zhu2015strong,Law2016direct,Santos2015laser,Zhang2018generation}, far beyond what can be achieved with permanent magnets. These magnetic fields, and the 10 to 100 kilotesla (100 MG to 1 GG) strengths envisioned for future experiments \cite{Korneev2015gigagauss}, may enhance inertial confinement fusion (ICF) \cite{Gotchev2009laser,Chang2011fusion,Slutz2012high}, accelerate particles more effectively \cite{Perez2013deflection,Arefiev2016enhanced}, and provide new capabilities for high-energy-density physics and laboratory astrophysics \cite{Matsuo2017magnetohydrodynamics,Santos2018laser}. Strong magnetic fields also open a new regime of laser-plasma interaction (LPI) physics \cite{Shi2018laser}, holding both the promise of useful new nonlinearities as well as the risk of damaging scattering in implosion-type experiments. 

A problem of particular importance in LPI physics is the scattering of laser beams from electron or ion waves. This can be either a deleterious effect, e.g.~the loss of energy from stimulated scattering or cross-beam energy transfer \cite{Froula2012increasing,Moody2012multistep,Marozas2018first}, or a beneficial instability useful for the construction of ultra-high-power plasma-based parametric amplifiers \cite{Malkin1999}. Plasma components in high-power lasers avoid damage thresholds set by solid-state optics, allowing extraordinarily-high-peak-power lasers to be envisioned without prohibitive beam diameters. Parametric plasma amplification, where a plasma wave compresses a long-duration high-energy pump into a short lower-frequency high-power seed, is a key component of proposed plasma-based laser amplifier chains. Current plasma amplifiers are based on two established mechanisms: stimulated Raman scattering (SRS) \cite{Malkin2000,Ping2004,Cheng2005,Ren2008,Yampolsky2008,Ping2009,
Vieux2011,Trines2011production,Turnbull2012,Depierreux2014,Toroker2014,Lehmann2014,Edwards2015,Edwards2017beam},
where a Langmuir wave provides a high growth rate but a large frequency downshift, or stimulated Brillouin scattering (SBS) \cite{Andreev2006,Lancia2010,Lehmann2012,Lehmann2013,Weber2013,Riconda2013,Guillaume2014demonstration,
Lehmann2016temperature,Lancia2016signatures,Chiaramello2016role,Edwards2016strongly,Jia2016distinguishing,Edwards2017xray}, where the ion-acoustic wave mediates the interaction, giving a lower growth rate but allowing higher efficiency energy extraction and a smaller frequency separation between seed and pump. The SBS frequency difference can be small enough to derive both from the same source \cite{Edwards2016short}. 

Here we demonstrate that the rich physics of strongly-magnetized plasma offers a third distinct mechanism for plasma amplification: scattering from the kinetic extensions of magnetohydrodynamic (MHD) waves, which exhibits an instability growth rate comparable to or higher than Raman scattering but a much smaller frequency downshift. Since this interaction disappears in unmagnetized plasma and lacks an analogue in non-ionized media after which it could be named, we will refer to it here as magnetized low-frequency (MLF) scattering for the mediating plasma response. MLF scattering is promising for amplification because 
(1) the growth rate is larger than Raman scattering, allowing rescaling for shorter interaction duration and plasma size, 
(2) the frequency of the plasma response, and thus the frequency difference between the pump and seed beams, is small, reducing the technical difficulties in creating the pump and seed together and leading to higher efficiencies,
(3) the interaction bandwidth is wide, supporting the generation of short-duration pulses,
and (4), group velocity dispersion in the magnetized medium is large, permitting self-compression of the broad-bandwidth amplified pulses to ultrashort durations. 

Although the role of magnetic fields in light scattering from plasmas has previously been studied, simulations have generally been restricted to external fields exactly parallel or perpendicular to the propagation vector ($\mathbf{k}$) of the driving laser field, where MLF scattering does not appear, and have focused on modifying Raman and Brillouin scattering \cite{Yin2013self,Paknezhad2016effect,Winjum2018quenching} or on upper-hybrid-wave mediation \cite{Shi2017laser,Jia2017kinetic}, which can be regarded as an improvement of Raman compression. In addition to an analytic model, we use particle-in-cell (PIC) simulations, allowing us to avoid the restrictions of a cold-fluid model, check for competing instabilities and unanticipated nonlinear or kinetic effects, and capture physics outside the three-wave-coupling model, which here is particularly important as multiple MLF branches may contribute to a single interaction. 

\begin{figure}
\centering
\includegraphics[width=\linewidth]{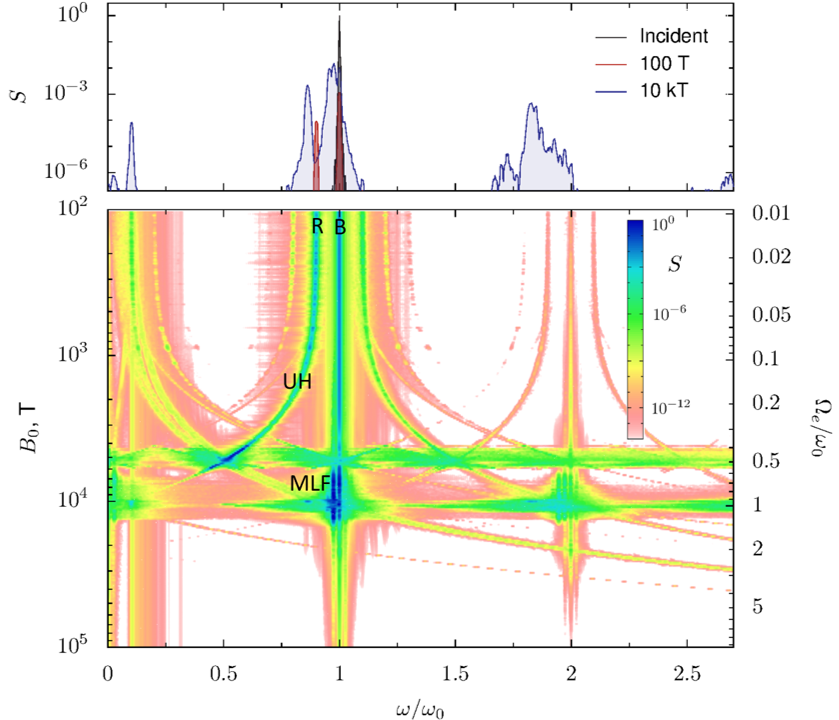}
\caption{Spectral energy distribution of light backscattered from magnetized plasma ($N = 0.01$, $T_e = T_i = 1$ eV) in PIC simulations. The magnetic field $B_0$ is varied between 100 T and 10 kT at $\theta=75^\circ$. The spectral energy is measured in vacuum ($\omega/\omega_0 = k/k_0$). The color scale indicates the spectral energy density logarithmically. (Top) The spectra of the incident beam and the reflected field at selected field strengths.}
\label{fig:spec}
\end{figure}

Consider spontaneous backscattering of a laser (frequency $\omega_0$) normally incident on an underdense ($\omega_0 < \omega_e$) semi-infinite homogeneous plasma, where $\omega_e = 4\pi n_e e^2/m_e$ is the plasma frequency. For negligible magnetic field magnitude ($B_0$), spontaneous Raman and Brillouin scattering may both occur. 
However, a much more complex picture for backscattering emerges when the electron cyclotron frequency ($\Omega_e = eB/m_e$) is not negligible compared to $\omega_0$ and $\mathbf{B}_0$ and $\bf{k}$ are neither exactly parallel nor perpendicular. 
Figure~\ref{fig:spec} presents the spectrum of backscattered light for $0.01 <\Omega_e/\omega_0 < 10$ and $\theta=\langle\mathbf{B}_0,\mathbf{k}\rangle=75^\circ$, as calculated with one-dimensional (1D) PIC simulations (EPOCH \cite{Arber2015contemporary}). With resolution $ \lambda / \Delta x = 50$ and 40 particles per cell, a linearly-polarized laser ($\mathbf{E} \perp {\mathbf{B}}_0$) with normalized field strength $a_0 = e E/m_e \omega_0 c = 0.01$ and wavelength $\lambda = 1$ $\upmu$m ($I = 1.38\times10^{14}$ W/cm$^2$) strikes an electron-proton plasma ($N = n_e/n_c = 0.01$ where $n_c$ is the critical density, temperature $T_e = T_i = 1$ eV). The spectral energy distributions for 300 simulations at different values of $B_0$ after 2 ps of interaction are provided. 
Backscattering for $0<B_0<100$ T does not vary significantly with $B_0$ on this scale. At $B_0 = 100$ T the spectrum is still dominated by Raman scattering at $\omega = \omega_0-\omega_e$ and Brillouin scattering at $\omega = \omega_0 - c_sk \approx \omega_0$, where $c_s$ is the sound speed. 

Above $B_0 = 1000$ T ($\Omega_e/\omega_0 > 0.1$) the spectral energy distribution changes substantially. When $\Omega_e$ is comparable to $\omega_e$ (here $\omega_e = 0.1\omega_0$), Raman scattering becomes scattering from the upper-hybrid (UH) wave, whose frequency shift is approximately $\omega_{\textrm{UH}} = \sqrt{\omega_e^2 + \Omega_e^2}$ for nearly perpendicular propagation; this can be seen as the curve in the Raman-UH line towards lower frequencies for larger magnetic fields. The transition to the upper-hybrid wave moderately improves laser amplification \cite{Shi2017laser,Jia2017kinetic}, but the most striking signature in Fig.~1 is the strong scattering that appears between $0.5 < \Omega_e/\omega_0 < 1.2$ at $\omega/\omega_0 \approx 0.99$. As described below, this results from scattering from kinetic MHD waves, i.e. MLF scattering, and is notable for both its large strength and small downshift from the pump frequency.

\begin{figure}
\centering
\includegraphics[width=\linewidth]{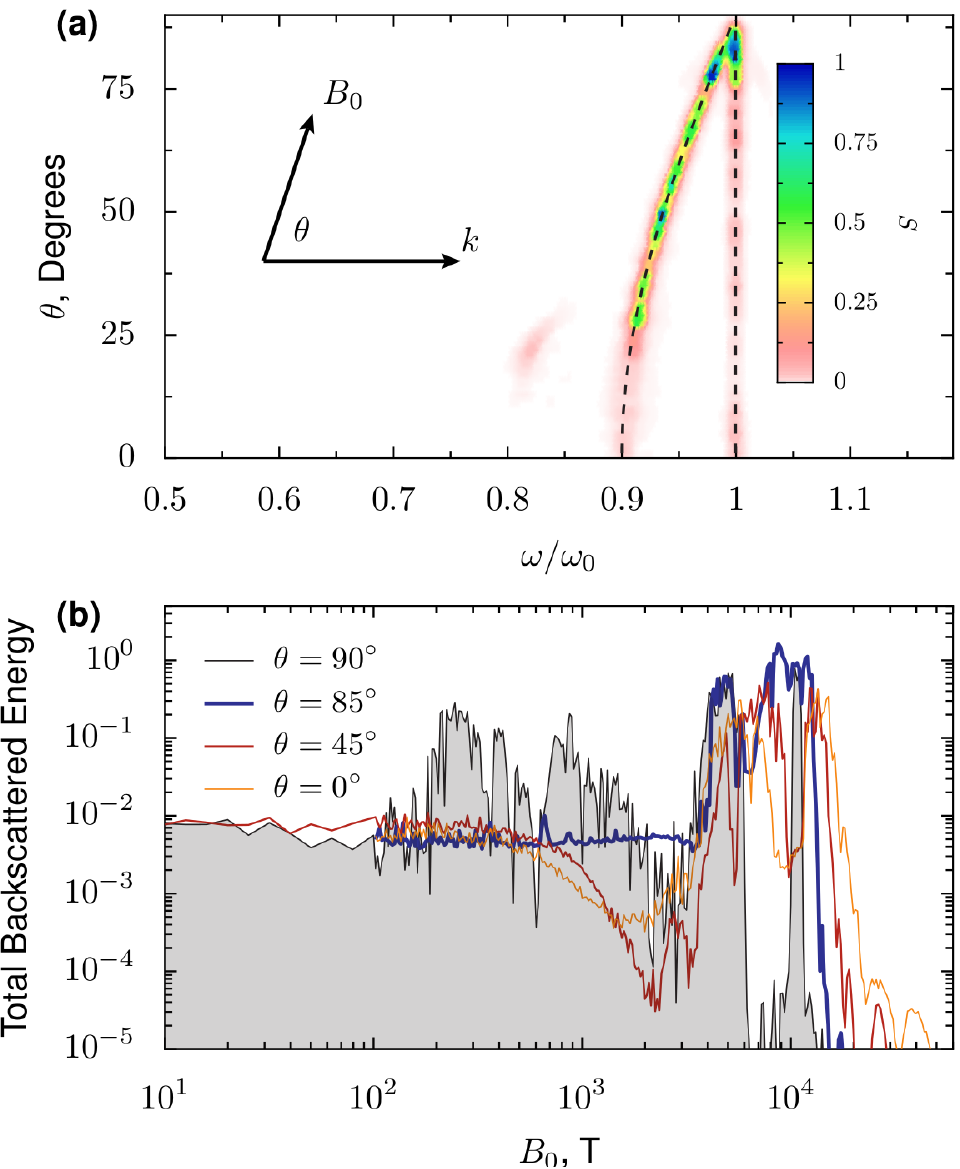}
\caption{PIC simulations of laser backscattering in magnetized plasma. (a) Backscattering spectrum from a magnetized electron-proton plasma as a function of angle at $\Omega_e/\omega_0 = 0.75$ ($B_0 = 8100$ T), illustrating dependence of MLF scattering on angle. Dashed lines show analytic predictions. (b) Total backscattered energy as a fraction of incident energy against magnetic field strength for selected angles.}
\label{fig:angle}
\end{figure}

Figure~2a provides the backscattered spectral energy distribution at $B_0 = 8100$ T ($\Omega_e/\omega_0 = 0.75$), where MLF scattering dominates. This scattering is strongest for oblique angles around 80$^\circ$ and goes toward 0 at $\theta = 90^\circ$. 
When $\theta \rightarrow 0$, the MLF scattering branch with the largest frequency downshift asymptotes to Raman scattering. 
The downshift frequency as a function of angle follows the analytic theory (dashed lines) and is relatively small for the large intermediate angles where the scattering is most powerful. 

A strong magnetic field also substantially affects the total scattered energy. As Fig.~2b illustrates, although previous work has shown suppressed backscattering for moderate applied fields (e.g.~$< 50$ T) \cite{Winjum2018quenching}, in the regime where $\omega_0$ and $\Omega_e$ are comparable, backscattering is dramatically enhanced. Due primarily to the contribution of MLF scattering, the strongest backscattering signal occurs at intermediate $\theta$ when $\Omega_e \approx \omega_0$. Interestingly, at even higher magnetic fields ($\Omega_e \gg \omega_0$), electron responses are suppressed and there is little backscattering for all $\theta$; although the required fields are large, this regime may ultimately prove useful for scattering suppression in compression experiments.

\begin{figure}
\centering 
\includegraphics[width=\linewidth]{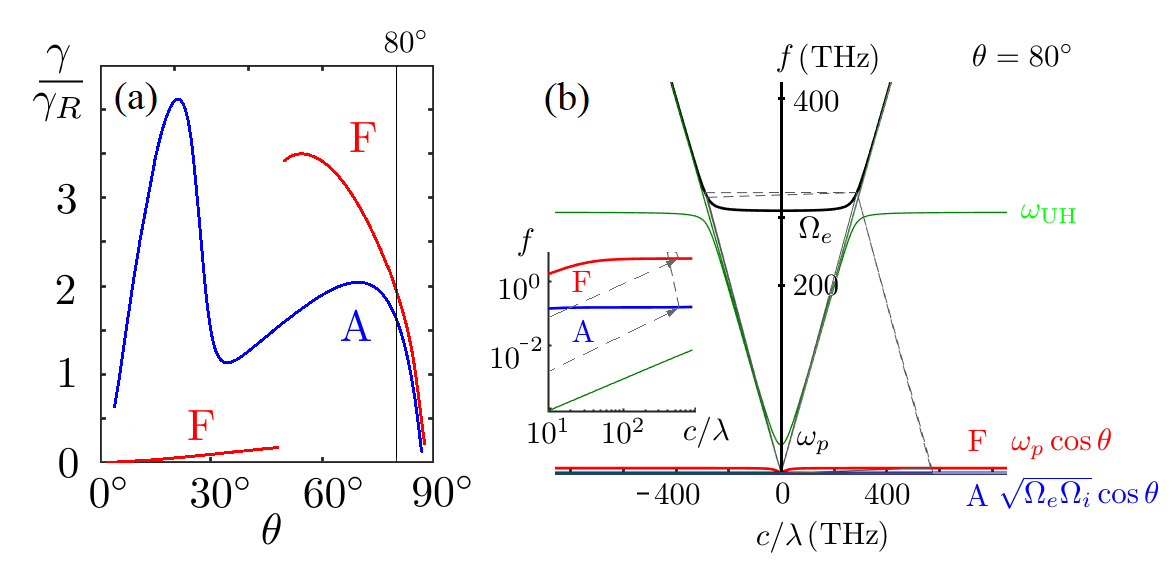}
\caption{Pulse compression mediated by high-$k$ asymptotes of MHD waves has significant advantages in the regime $\omega\gtrsim\Omega_e\gg\omega_p$. First, the linear growth rate $\gamma$ of MHD-wave mediation (a) is comparable or even larger than the Raman growth rate $\gamma_R$ at the same plasma density, and the frequency downshift is similar to that of Brillouin. Second, the bandwidth of MHD-wave mediation is large at oblique angle $\theta=\langle\mathbf{B}_0,\mathbf{k}\rangle$, because the fast MHD wave (F, red) and the shear Alfv\'en wave (A, blue) have overlapping bands. Third, electromagnetic waves in magnetized plasma have smaller group velocity $v_g=\partial\omega/\partial k$ and larger group velocity dispersion $G=\partial^2k/\partial\omega^2$, as can be seem from the wave dispersion relation at $\theta=80^\circ$ (b), corresponding to the vertical line in (a). Plasma parameters here are identical to those in Fig.~\ref{fig:pic} where resonant slow-MHD-wave mediation, which is viable in general, is forbidden by the three-wave resonance conditions. The resonance conditions produce a jump in $\gamma$ for F mediation near $\theta\approx56^\circ$. Inset (b): the dispersion relation for low frequency waves. The gray dashed lines delineate the resonance matching conditions.}
\label{fig:analytic}
\end{figure}

The MLF instability can be predicted analytically by a two-fluid model \cite{Shi2017three}. In magnetized two-species plasmas, the low-frequency waves contain three branches. In the low-$k$ limit, these branches are known as the MHD waves, namely, the fast (compressional) MHD wave, the shear Alfv\'en wave, and the slow (sound) MHD wave. In the high-$k$ limit, these wave branches continue beyond the validity of MHD theory. Depending on $\theta$, the species mass ratio, and the ratio of $c_s$ and the Alfv\'en speed $v_A$, the three branches cross and hybridize in a variety of ways \cite{Stringer1963low}. When the laser propagates parallel to the magnetic field under the conditions we consider here, specifically $\Omega_e\gg\omega_e$ and $v_A\gg c_s$, the upper branch is the Langmuir wave, the lower branch is the ion cyclotron wave, and the bottom branch is the sound wave. When the propagation is nearly perpendicular to $\mathbf{B}_0$, the upper branch is the lower-hybrid wave and the frequencies of the lower and bottom branches, which are both proportional to $\cos\theta$, asymptote to zero. At general angles, these waves have intermediate frequencies between the limiting cases. Furthermore, at wavevector $2k_0$, these waves are mostly longitudinal and the three branches have not yet crossed over. 

A particularly interesting regime for laser pulse compression occurs when $\omega_0$ lies slightly above $\Omega_e$ with $\Omega_e \gg \omega_e, c_s k$. For these conditions the plasma is strongly magnetized, and the right-hand component of the laser resonantly excites electron gyromotion. When $\mathbf{k}$ is parallel or perpendicular to $\mathbf{B}_0$, laser absorption by a single electron is forbidden by energy-momentum conservation. However, at an oblique angle resonant absorption is possible for $\omega_0\gtrsim\Omega_e$. The resonantly driven electron-cyclotron motion produces an enhanced diamagnetic field. In this screened magnetic field, the gyrofrequency of electrons is reduced. Upon deexcitation, the electrons emit an electromagnetic wave with slightly downshifted frequency, and the magnetic field returns to its original strength. This scattering process relies on the resonance with the cyclotron motion of electrons instead of the resonances provided by quasi particles. Therefore, electron-mediated MLF scattering has a larger cross section than Raman scattering, which relies on plasmons as the mediating quasi particles, as well as Brillouin scattering, which relies on phonons. Since the plasma wavevector is comparable to the laser wavevector $k_0\gg\Omega_i/c$, the mediating low-frequency waves are the high-$k$ limits of MHD waves \cite{Stringer1963low}. At intermediate $\theta$, the MHD-mediated parametric growth rates \cite{Shi2017three} are comparable to the Raman growth rate in an unmagnetized plasma of the same density (Fig.~\ref{fig:analytic}a).

\begin{figure}
\centering
\includegraphics[width=\linewidth]{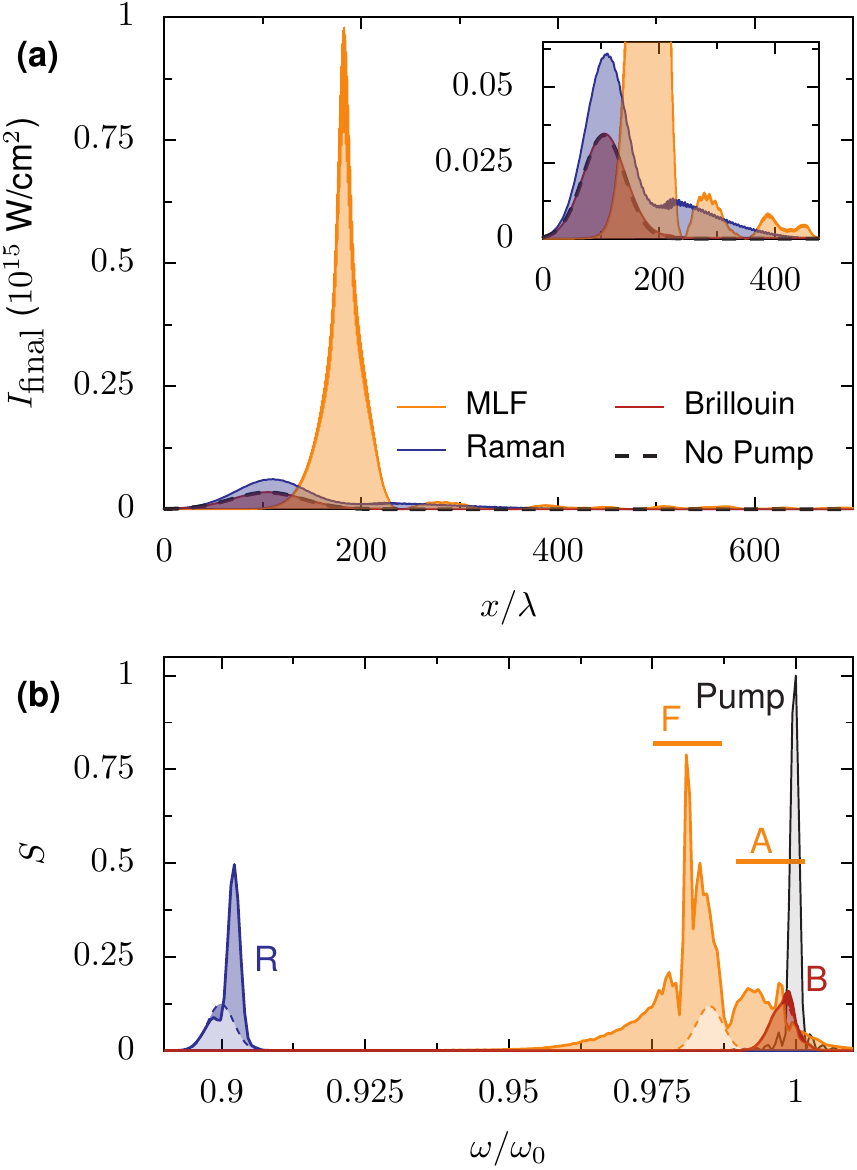}
\caption{PIC simulations of parametric amplification due to stimulated Raman, Brillouin, and MLF scattering. (a) Final amplified seed after interaction in a plasma ($N=0.01$, $L=175\lambda$) with a counterpropagating pump ($a_0 = 0.007$, $\lambda = 1$ $\upmu$m, $I = 6.7 \times 10^{13}$ W/cm$^2$). (b) Spectra of the initial (dashed line) and amplified (solid line) seed pulses for each coupling mechanism, together with the pump spectrum. For all simulations, $\lambda/\Delta x = 150$, 200 particles/cell, $T_e = T_i = 1$ eV and $m_e/m_i = 1836$. For MLF $B_0 = 9900$ T and $\theta = 80^\circ$. The initial seed intensities are half the pump intensity.}
\label{fig:pic}
\end{figure}

To examine whether the MLF instability is useful for laser amplification we conducted PIC simulations with a counterpropagating pump-seed geometry. Figure~\ref{fig:pic} compares SRS, SBS, and stimulated MLF scattering. In each case the plasma density ($N = 0.01$), plasma length ($L = 175$ $\upmu$m $ = 175 \lambda$) and pump amplitude ($a_0 = 0.007$) were the same. The initial seed duration was 400 fs and the pump was uniform in time. In the magnetized case, both the seed and pump were elliptically polarized with Stokes parameters $Q = 0.78I_0,~U = 0.49I_0,~V = 0.38I_0$ for propagation through magnetized plasma. In the unmagnetized case, linear polarization was used. For each mechanism, simulations were conducted at varied seed wavelength, with the strongest response presented. In comparison to SRS and SBS, MLF scattering exhibits a much higher growth rate and broader spectral response; the amplified pulse both contains more energy and is shorter. The simulations suggest that high-power, few-cycle pulses could be generated with relatively short plasma lengths. The frequency downshift is small enough and the spectral response sufficiently broad that the pump and seed could be prepared from the same source.

Of interest in Fig.~\ref{fig:pic}b is the broad bandwidth generated by MLF scattering, which supports ultrashort pulses.  
The bandwidth is broad for two reasons. First, the high growth rate of each individual MHD wave produces a large bandwidth. Second, multiple MHD waves, in this case the fast and Alfv\'en waves, may be simultaneously excited because the bandwidth of the sub-picosecond seed pulse is comparable to the frequencies of MHD waves. The overlapping MHD bands then produce an even larger total bandwidth for MLF scattering.
For these parameters, a particularly strong interaction is mediated by the longitudinal electron oscillation with frequency \mbox{$\sim\omega_p\cos\theta$} (Fig.~\ref{fig:analytic}, red), which corresponds to peaks near $\omega/\omega_0\approx0.98$ in Fig.~\ref{fig:pic}b (orange, F). This wave, with both magnetic and density perturbations, carries the recoil energy and momentum after electrons resonantly absorb and then emit the laser photons. 
The magnetic perturbation is due to the oscillating diamagnetic screening, and the density perturbation is present because the electrons cannot move quickly enough to neutralize the space charge when $k_0>\omega_p/c$. 
This mediating wave is the fast MHD wave in the large-$k$ limit, which has a very small group velocity when the plasma is cold ($c_s\ll v_A$). In addition to the fast MHD wave, the large-$k$ limit of the shear Alfv\'en wave with frequency $\sim \sqrt{\Omega_e\Omega_i}\cos\theta$ also provides coupling (Fig.~\ref{fig:analytic}, blue). The Alfv\'en-wave mediation is responsible for the peak near $\omega/\omega_0\approx0.995$ in \mbox{Fig.~\ref{fig:pic}b (orange, A)}. The slow MHD wave in general also provides large coupling, but for these specific parameters the three-wave resonance conditions cannot be satisfied. In short, PIC calculations show that MLF-mediated scattering produces a large growth rate over a wide bandwidth and a small frequency downshift.

One further advantage for pulse compression exists in magnetized plasma due to the increased group velocity dispersion, which can in principle be exploited by chirping the seed pulse. For Raman amplification, shorter pulse durations require higher plasma density to increase the plasma frequency, but this increases detrimental secondary effects, including forward scattering, collisional damping, and the modulational instability. Chirping the seed pulse \cite{Toroker2012seed} has been proposed for avoiding these effects by reducing the plasma length. Since lower-frequency electromagnetic waves propagate at smaller group velocity than higher-frequency waves, a chirped seed pulse with \mbox{$\partial\omega/\partial t>0$} will self-contract if the plasma is terminated when the seed pulse reaches its shortest duration. Under the strongly-magnetized conditions considered here, where group velocity dispersion is larger than for unmagnetized plasma, chirping offers a substantial advantage \cite{Toroker2012seed}. The equation governing the evolution of the pulse envelope $b$ can be written $(\partial_t+v_g\partial_x+i\tau\partial_t^2)b=0$ in the absence of a pump and the modulational instability. The group velocity dispersion coefficient $\tau=-\frac{1}{2}G v_g$, where $G=\partial^2k/\partial\omega^2$. Using the dispersion relation of the extraordinary wave, $\tau$ can be evaluated as a function of $\omega$. For strongly-magnetized plasma, $\tau$ is almost three orders of magnitude larger than its unmagnetized value and, consequently, a chirped seed pulse can be rapidly compressed in a much shorter plasma length.

It is worth noting that the dependence of the frequency downshift on magnetic field strength rather than density may reduce the sensitivity of the process to plasma inhomogeneities. This prevents using density gradients for precursor suppression \cite{Tsidulko2002suppression}, but matching pump chirping \cite{Malkin2000detuned} to the field gradient in a diverging magnetic field could serve a similar role. 

Although the magnetic fields discussed here are high, the 5-10 kT (50-100 MG) strengths required are near current state-of-the-art capabilities. Furthermore, scaling to longer wavelengths reduces the absolute field requirements; the physics are governed by $\Omega_e/\omega_0$. This means, for example, that the regime of interest is reached above 500 T for CO$_2$ lasers ($\lambda = 10.6$ $\upmu$m). The counter-propagating geometry required for amplification is relatively complex, but signatures of spontaneously scattering may be visible for more readily achieved experimental conditions, and consideration of this enhanced scattering mechanism may be necessary to avoid damage to large laser-compression facilities. 

In conclusion, we have demonstrated that a laser will be strongly backscattered from kinetic MHD waves when it propagates in a plasma at an oblique angle to a sufficiently-strong magnetic field. 
This powerful backscattering is potentially useful for laser amplification because the compression and amplification occur more quickly than for Raman amplification at the same density, and the frequency downshift, at less than 1\%, is comparable to that of Brillouin amplification. Additionally, substantial spectral broadening is observed in PIC simulations, supporting amplified pulses with ultrashort durations. 
MHD-wave-mediated instabilities represent a new class of backscattering mechanisms, distinct from both Raman and Brillouin scattering, which may be critical in high-intensity laser experiments, including ICF, with strong applied or self-generated magnetic fields. 

\begin{acknowledgments}
This work was supported by NNSA Grant No.~DENA0002948, AFOSR Grant No.~FA9550-15-1-0391, NSF Grant No.~PHY 1506372, and DOE grant No.~DE-SC0017907. Computing support for this work came from the High Performance Computing Center at Princeton University. The EPOCH code was developed as part of the UK EPSRC 300 360 funded project EP/G054940/1.
\end{acknowledgments}

\end{document}